\documentclass[10pt,aps,amsmath,showpacs,amssymb,nofootinbib,pre,twocolumn]{revtex4}
\usepackage[dvips]{graphicx}
\usepackage[dvips,usenames]{color}
\usepackage{amsmath}

\newcommand{\av}[1]{\langle {#1} \rangle}

\begin{document}

\title{Localization transition, Lifschitz tails and rare-region effects 
in network models} 

\author{G\'eza \'Odor}
\affiliation{Research Center for Natural Sciences, 
Hungarian Academy of Sciences} 
\address{P. O. Box 49, H-1525 Budapest, Hungary}

\date{\today}

%%%%%%%%%%%%%%%%%%%%%%%%%%%%%%%%%%%%%%%%%%%%%%%%%%%%%%%%%%%%%%%%%%%%%%%%%
\begin{abstract}
%%%%%%%%%%%%%%%%%%%%%%%%%%%%%%%%%%%%%%%%%%%%%%%%%%%%%%%%%%%%%%%%%%%%%%%%%
Effects of heterogeneity in the suspected-infected-susceptible model 
on networks are investigated using quenched mean-field theory. 
The emergence of localization is described by the distributions of the 
inverse participation ratio and compared with the rare-region effects 
appearing in simulations and in the Lifschitz tails. 
The latter, in the linear approximation, is related to the spectral density 
of the Laplacian matrix and to the time dependent order parameter. 
I show that these approximations indicate correctly Griffiths Phases both 
on regular one-dimensional lattices and on small world networks 
exhibiting purely topological disorder. I discuss the localization 
transition that occurs on scale-free networks at $\gamma=3$ degree
exponent.
\end{abstract}

\maketitle

%%%%%%%%%%%%%%%%%%%%%%%%%%%%%%%%%%%%%%%%%%%%%%%%%%%%%%%%%%%%%%%%%%%%%%%%%
\section{Introduction}
%%%%%%%%%%%%%%%%%%%%%%%%%%%%%%%%%%%%%%%%%%%%%%%%%%%%%%%%%%%%%%%%%%%%%%%%%

Epidemic spreading in complex networks such as biological populations and
computer networks is of great interest, both for practical applications and
from a fundamental point of view \cite{ABR02,DFR02,NR10}.
Simple models, like the Contact Process (CP) \cite{harris74,liggett1985ips} 
has been introduced and studied intensively by various techniques. 
They can also be considered as simple models of information spreading in 
social \cite{pv04} or in brain networks \cite{MMNat}.
In these models sites can be infected (active) or susceptible (inactive).
Infected sites propagate the epidemic to all of their neighbors,
with rate $\lambda$, or recover (spontaneously deactivate) with rate $\nu=1$.
The susceptible-infected-susceptible (SIS) \cite{SIS} model differs slightly 
from the CP, in which the branching rate is normalized by $k$,
the number of outgoing edges of a vertex permitting an analytic
treatment via symmetric matrices.
By decreasing the infection (communication) rate of the neighbors
a continuous phase transition may occur at some $\lambda_c$ critical
point from a steady state with finite activity density $\rho$ to an
inactive one, with $\rho=0$. The latter is also called absorbing, 
since no spontaneous activation of sites is allowed. In case of the
SIS $\lambda_c=0$ in networks with a degree distribution decaying 
slower than an exponential \cite{BCS13} \footnote{Note, that some
recent studies debate this, see: \cite{DGMS03,LSN12,Ferrnew}}.
The transition type is continuous and belongs to the directed 
percolation universality class \cite{DickMar,rmp,odorbook,HH}.

In real systems various heterogeneities occur, that may cause 
deviations from the results of the homogeneous models. From the
homogeneous system point of view if the disorder varies rapidly
both in space and time, its contribution can be described by an
increased temperature or noise of the system \cite{odorbook}. 
In the quasi-static limit, when the variation of the heterogeneity
is much slower the dynamics of the pure model we can consider 
it as a quenched disorder. It causes a memory effect, whose 
relevancy has been studied in quantum and nonequilibrium systems
(see \cite{Vojta}).

In networks, with finite topological dimension, defined as $N\propto r^D$, 
where $N$ is the number of nodes within the (chemical) distance $r$, 
it was shown  \cite{PhysRevLett.105.128701}, that disorder can be 
relevant. Heterogeneities can induce arbitrarily large, rare-regions (RR), 
changing their state exponentially slowly as the function of their sizes, 
induce so called Griffiths Phases (GP) \cite{Griffiths,Vojta}. 
In these phases the dynamics is slow and non-universal and at the
phase transition point it is even slower, logarithmic dynamical scaling
may occur. These heterogeneities can be explicit features of the
interactions or maybe the result of the topology of the graph.

Recent observations show generically slow time evolution in various system.
For example in the working memory of the brain \cite{Johnson}
or in recovery processes following virus pandemics \cite{pv04,PV01,KK11} 
power-law type of time dependencies have been found, resembling of 
dynamical critical phenomena \cite{Chialvo}. 
In social networks the occurrence of generic slow dynamics was suggested 
to be the result of the non-Markovian, bursty behavior of agents 
in small world networks \cite{KK11}.
Very recently it has been shown \cite{burstcikk}, that bursty dynamics 
can arise naturally, in network models as the consequence of power-law 
decaying auto-correlations due to the {\it collective behavior of 
Markovian variables.}

Disorder effects are stronger in quantum systems, where the
thermal noise does not fade effects of the quenched noise. 
However, in several cases the critical behavior is dictated by an
infinitely strong disorder fixed point, resulting in robust
universality classes, that can be observed even in classical models. 
In particular, the same universal behavior occurs in disordered
quantum Ising chains and the CP \cite{IMrev}.  
The dynamics of the CP far in the absorbing state, can be mapped to 
the quantum-mechanical one, described by the disordered Hamiltonian 
of the Anderson type (see \cite{KK93}). 

Heterogeneous Mean Field (HMF) theory provides a good approximation in 
network models, when the fluctuations are irrelevant 
\cite{boguna09:_langev,FCP12,Fer14}. 
To describe quenched disorder in networks the so-called 
Quenched Mean-Field (QMF) approximation is introduced 
\cite{CWW08,Mieg09,CP10,GDOM12} and heterogeneities of the 
steady state are quantified by calculating the Inverse Participation 
Ratio (IPR) of the principal eigenvector of the adjacency matrix.
Effects of the quenched disorder on the dynamical behavior of SIS have
recently been compared using QMF approximations in different network models.
Numerical evidences have been provided for the relation of {\it localization 
to RR effects, that slows the dynamics \cite{wbacikk,basiscikk}.}

The success of this relation is the consequence of the fact that GP effects 
arise even in the active phase, where localization of the steady state 
can be traced by the IPR value. 
Although for best understanding the effects of dynamical fluctuations
should also be taken into account, such approaches, like renormalization 
group methods (RG) \cite{Monthus,KI11,JK13} have some limitations. 
For example, strong disorder RG works around an infinite disorder 
fixed point, which is not always present, still Griffiths singularities 
can co-exist with the clean critical behavior \cite{HoyosVojta}. 
Furthermore, this method cannot handle models with pure topological 
inhomogeneity. 

In this work I show that the QMF theory describes localization in the
one-dimensional SIS model, with quenched disorder, in agreement with the 
expectation that RR effects and GP should occur below the critical point. 
I extend previous localization studies by considering distributions of
the IPR and eigenvalues, casting more light on the localization transition
of SIS in various complex networks.
In particular, I investigate SIS on scale-free (SF) networks, possessing 
$P(k)\propto k^{-\gamma}$ degree distributions and provide numerical
evidence for a localization transition at $\gamma=3$.

%%Lif intro
Very recently Moretti and Mu\~noz \cite{MMNat} have investigated hierarchical,
brain networks by simulations and QMF approximations. They gave a brief 
overview about the relation of slow dynamics and Lifschitz tails 
in synchronization and spreading models \cite{Matj}. 
Lifschitz tails have provided valuable information in regular, 
equilibrium systems about the Griffiths singularities 
(see for example \cite{Nus91}). In network models they have been 
studied in mathematics literature mainly \cite{KKM}. 
In graph theory there is a growing interest 
in spectral properties of linear operators, mostly of the adjacency
matrix or the graph Laplacian (see for example \cite{Chung,Miegbook}). 
In physics literature Samukhin et al. \cite{SDM08} provided analytical 
forms for the Laplacian spectrum of complex random networks and for the
dynamical two-point functions of random walks running on them. 
They pointed out that the minimum degree of vertexes, is important
for the dynamics, which is related to the lower edge of the Laplacian.
On the other hand numerical evidences have been shown that the 
spectral gaps at the lower edge describe well the slow-down of dynamics
due to disorder in models like the CP \cite{TLNG05} or by synchronization
transition \cite{BC,DNM06}.
This is based on the validity of linearization near the phase transition
point. In this case the probability distribution at the lower tail of the 
Laplacian can be considered the density of states, the Lifschitz tail of
the disordered network model. If it holds it enables us to describe the 
dynamics near the critical point.
In this study I calculate the lower tail distributions of the Laplacian
of the networks considered and test how well does it describe the 
GP behavior of the SIS.

%%%%%%%%%%%%%%%%%%%%%%%%%%%%%%%%%%%%%%%%%%%%%%%%%%%%%%%%%%%%%%%%%%%%%%%%%
\section{Summary of earlier studies: localization versus RR effects}
%%%%%%%%%%%%%%%%%%%%%%%%%%%%%%%%%%%%%%%%%%%%%%%%%%%%%%%%%%%%%%%%%%%%%%%%%

Starting from the master equation for state vectors of site occupancies,
$\vert P_{ ( n_1,n_2,...,n_N ) } (t)\rangle$
where $n_i=0$ or $1$, one can derive the QMF theory for the SIS model 
\cite{Mieg09,GDOM12}. 
Although QMF neglects the dynamical correlations, it can take into account 
heterogeneities of the network by considering the vector of infection 
probabilities $\rho_i(t)$ of node $i$ at time $t$
\begin{equation}
\label{qmfsis}
\frac{d\rho_i(t)}{dt} = -\rho_i(t) + \lambda (1-\rho_i(t))\sum_{j=1}^N A_{ij}w_{ij} \rho_j(t)~.
\end{equation}
Here $A_{ij}$ is an element of the adjacency matrix and $w_{ij}$ describes 
the possibility of weights attributed to the edges.
For large times the SIS model evolves into a steady state, with an order
parameter $\rho \equiv \av{\rho_{i}}$.
This equation with $i \leftrightarrow j$ symmetric weights can be 
treated by a spectral decomposition on an orthonormal eigenvector basis. 
Furthermore the non-negativity of
the $B_{ij} \equiv A_{ij}w_{ij}$ matrix involves a unique, real, non-negative 
largest eigenvalue $y_{M}$. 

For $t\rightarrow \infty$ the system evolves into a steady state and the
infection probabilities can be expressed via $B_{ij}$ as
\begin{equation}
\rho_{i}=
\frac{\lambda\sum_{j}B_{ij}\rho_{j}}{1+\lambda\sum_{j}B_{ij}\rho_{j}} \ .
\label{SIS2}
\end{equation}
The order parameter (prevalence) $\rho \equiv \av{\rho_{i}}$ becomes
finite above an epidemic threshold $\lambda_{c}$.
In the QMF approximation one finds $\lambda_c$ and $\rho(\lambda)$ 
around it from the principal eigenvector.
Using a Taylor expansion of $\rho$ one can solve Eq.~(\ref{SIS2}) 
and find that the threshold is related to the largest eigenvalue of 
$B_{ij}$ as: $1/\lambda_c=y_{M}$.
The order parameter near, above $\lambda_c$ can be approximated via
\begin{equation} \label{expansion}
\rho(\lambda) \approx a_1 \Delta + a_2 \Delta^2 + ... \ ,
\end{equation}
where $\Delta = \lambda y_{M}{-}1 {\ll} 1$ and the coefficients
\begin{equation}
a_j = \sum_{i=1}^N e_i(y_j)/[N \sum_{i=1}^N e_i^3 (y_j)] \,
\label{epsilon}
\end{equation}
are functions of eigenvectors ${\bf e}(y_j)$ of the largest eigenvalues
($j=M, M-1, M-2, ...$) of $B_{ij}$.
This expression is exact, if there is a gap between $y_M$ and $y_{M-1}$ 
\cite{Mieg12}.

It was proposed in \cite{GDOM12} and tested on weighted Barabasi-Albert 
models \cite{wbacikk} that the localization of activity in the active steady
state can be characterized by the $IPR$ value, related to the eigenvector of 
the largest eigenvalue $\mbox{\boldmath$e$}(y_M)$ as
\begin{equation} \label{IPR}
I(N) \equiv \sum_{i=1}^{N} e_{i}^{4}(y_M) 
\end{equation}
This quantity disappears as $\sim 1/N$ in case of homogeneous eigenvector 
components or remains finite if the activity is concentrated 
on a finite number of nodes.

%%%%%%%%%%%%%%%%%%%%%%%%%%%%%%%%%%%%%%%%%%%%%%%%%%%%%%%%%%%%%%%%%%%%%%%%%
\section{Lifschitz tails in network models}
%%%%%%%%%%%%%%%%%%%%%%%%%%%%%%%%%%%%%%%%%%%%%%%%%%%%%%%%%%%%%%%%%%%%%%%%%

Besides IPR calculation, that works in the active steady state, some other 
way to check RR effects would be desirable. The study the spectrum tail 
of the Laplacian will be introduced here in the hope of providing
information about GPs below the critical point of the SIS.
The Laplacian matrix of a graph is defined as
\begin{equation}\label{Lapl}
L_{ij}=\delta_{ij}\sum_l A_{jl} - A_{ij} \  ,
\end{equation}
which takes values $-1$ for pairs of connected vertexes and the degree
$k_i$ in the diagonal. The Laplacian is positive-semi-definite, i.e.:
$\Lambda_i \ge 0$ and $\Lambda_1 =0$.
The smallest non-zero eigenvalue $\Lambda_2$ is called the spectral gap.

Near the critical point, in the inactive phase we can linearize
the dynamical equation of SIS (\ref{qmfsis}) as
\begin{equation}\label{eqL}
\frac{d\rho_i(t)}{dt} = -\rho_i(t) + \lambda \sum_j B_{ij} \rho_j(t) \ .
\end{equation}
We can rewrite it, using the weighted (symmetric) Laplacian matrix
\cite{MO08,M11}
\begin{equation}
L_{ij}=\delta_{ij}\sum_l B_{jl} - B_{ij} \ ,
\end{equation}
which has the sums of weights in the diagonal, expressed
by the Kronecker delta ($\delta_{ij}$), as follows
\begin{equation}
\frac{d\rho_i(t)}{dt}= \left[ \lambda \delta_{ij}\sum_l B_{jl} - 1 \right] \rho_i(t) 
- \lambda \sum_j  L_{ij}\rho_j(t) \ .
\end{equation}
A linear stability analysis can be performed above the critical point, 
similarly to the synchronization process \cite{BC}.
For the normal modes of the perturbations above the absorbing state we can write
\begin{equation}
\frac{d \rho_i(t)}{dt} = - \lambda \sum_j L_{ij} \rho_j(t) \ .
\end{equation}
By this approximation we replaced the diagonal elements in Eq.~(\ref{eqL}),
from $-1$ to $-\lambda L_{ii}$, which increases the spontaneous recovery 
rate $\nu$ of sites, pushing the system deeper into the inactive phase.
In spreading models it is known that the value of $\nu$ can modify non-universal 
quantities, shift $\lambda_c$, but in the inactive phase, where this approach 
is applied, it is not expected to induce relevant RR effects, it can make 
them weaker and harder to detect. However, I have confirmed this 
approximation in case of CP on networks with purely topological disorder.

Using the spectrum of $L_{ij}$ one can make the eigenvalue expansion
\begin{equation}
\rho_i(t) = \sum_{jl} e^{-\lambda\Lambda_l t} 
f_i(\Lambda_l) f_j(\Lambda_l) \rho_j(0) \ , 
\end{equation}
where $f_i(\Lambda_l)$ is $i$-th the component of the $l$-th eigenvector of
the Laplacian.
The total density is determined by the lowest eigenvalues of the spectrum
\begin{equation}
\rho (t) \sim \sum_{l=2}^N e^{- \lambda \Lambda_l t}
\end{equation}
for any network. In finite systems there is always a finite $\Lambda_2 > 0$ 
gap, causing exponential cutoff in the decay of the order parameter. 
In this study I consider $P(\Lambda)$ above $\Lambda_2$, i.e. shift 
the numerically obtained distributions to zero and express $\rho(t)$ 
as the Laplace transform of $P(\Lambda$) in the continuum limit
\begin{equation}
\rho(t) \propto \int_{\Lambda_2}^{\Lambda_M} d\Lambda \ 
P(\Lambda) e^{-\lambda\Lambda t} \ ,
\end{equation}
where $\Lambda_M$ corresponds to the experimentally determined
end of tail value of the finite network.
Note, that the control parameter $\lambda$ appears as a constant,
which can induce non-universal power-laws in the inactive GP.

One can also take into account the original diagonal elements of (\ref{eqL}), 
if one considers the CP instead of SIS, where the interactions are normalized 
by the degree as $\lambda_i/k_i$. In case of purely topological 
heterogeneities the linearized, governing equation takes the form:
\begin{equation}
\label{qmfcp}
\frac{d\rho_i(t)}{dt} = 
-\rho_i(t) + \sum_{j=1}^N \frac{\lambda}{k_j} A_{ij}\rho_j(t) \ .
\end{equation}
thus the sum of non-diagonal elements: $\frac{\lambda}{k_j} A_{ij}$ is 
constant: $\lambda$.
The eigenvalue spectrum of the matrix $L'_{ij} = \delta_{ij} - \frac{\lambda}{k_j} A_{ij}$
is the linear combination of the normalized Laplacian:
$L'_{ij} = \frac{\lambda}{k_j} L_{ij} - \delta_{ij}( \lambda-1)$ for such models. 
Therefore, by performing a spectral analysis of $L'_{ij}$ we can
investigate the lower gap behavior. The penalty is that we have
non-symmetric matrices, which can be diagonalized by slower algorithms.
I have determined this spectrum for uncorrelated random and
generalized small networks (for definition see later sections)
and found tails very similar as that of SIS, except from the 
linear transformation.

%ER Lifschitz test

\begin{figure}
\includegraphics[height=5.5cm]{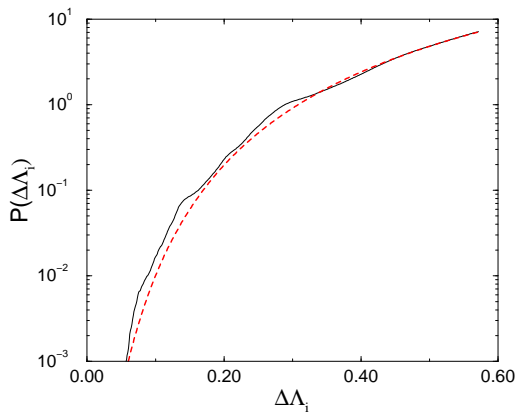}
\caption{\label{LER4}(Color online) Lifschitz tail of the ER graph
with $\langle k\rangle =4$ and $N=10^4$. The dashed line shows a numerical
fit with the form (\ref{PER}) as: 
$2400 \Delta\Lambda_i^{1/10}\exp(-4.5/(\Delta\Lambda_i^{1/2}). $}
\end{figure}

For comparison I calculated the Laplacian eigenvalue spectrum of the 
Erd\H os R\'enyi (ER) \cite{ER} graph with $N=10^4$ nodes and 
$\langle k\rangle =4$ average degree.
Averaging over $2.5\times 10^5$ random graph realizations and histogramming,
with the bin size $\delta\Lambda=0.001$ one can determine numerically
the probability distribution $P(\Lambda_i)$ in the region $0 < \Lambda < 0.6$.
The gap size due to the finite system was $\Lambda_2=0.036$, that
I subtracted: $\Delta\Lambda_i=\Lambda_i-\Lambda_2$.
A good fitting can be obtained with the cumulative distribution derived 
form \cite{SDM08} with the numerical factors shown in Fig.~\ref{LER4}.
%(\ref{pER})
\begin{equation}\label{PER}
P(\Delta\Lambda) \simeq \Delta\Lambda^{1/10} e^{-a/\sqrt{\Delta\Lambda}} \ . 
\end{equation}
%%%%%%%%%%%%%%%%%%%%%%%%%%%%%%%%%%%%%%%%%%%%%%%%%%%%%%%%%%%%%%%%%%%%%%%%%
%%%%%%%%%%%%%%%%%%%%%%%%%%%%%%%%%%%%%%%%%%%%%%%%%%%%%%%%%%%%%%%%%%%%%%%%
The Laplace transform of (\ref{PER}) predicts the long time 
asymptotic behavior of the density decay
\begin{equation}
\rho(t) \sim e^{ - (c t \lambda)^{1/3}} %t^{-y} \ ,
\end{equation}
which is a $\lambda$ dependent stretched exponential time dependence.
Numerical simulations of the disordered CP on ER graphs have obtained
indeed $\lambda$ dependent stretched exponential density decay behavior
below the critical point \cite{GPCNlong}.
However, the validity of Eq.~(\ref{PER}) is limited to 
$t^{1/3}/\ln N << 1$ \cite{SDM08}, hence in numerically accessible 
systems this should be observable for very short times only. 
In density decay simulations of SIS on pure ER systems with 
$N=5\times 10^6$ the effect of topological disorder can be seen
for very early times, otherwise exponential decay is observed.

Contrary, for a power-law distributed $P(\Lambda)$ the Laplace
transformation results in a power-law decaying density
\begin{equation}
\rho(t) \propto \int_{\Lambda_2}^{\Lambda_M} d\Lambda \ \Lambda^x 
e^{-\lambda\Lambda t}
\propto t^{-\lambda(x+1)} \ ,
\end{equation}
which suggests a GP behavior for the model.
Therefore, in the following sections I determine numerically
$P(\Lambda)$ for certain models and determine how well can the
tail behavior be fitted by a power-law form.
By knowing dynamical simulation results about the existence of GPs
in these systems I test the predicting power of this approach.
Later, I apply the method to more difficult cases and try to 
support statements about existence of GPs in them.

%%%%%%%%%%%%%%%%%%%%%%%%%%%%%%%%%%%%%%%%%%%%%%%%%%%%%%%%%%%%%%%%%%%%%%%%
\section{QMF of the one-dimensional SIS model with quenched infection rates} 
%%%%%%%%%%%%%%%%%%%%%%%%%%%%%%%%%%%%%%%%%%%%%%%%%%%%%%%%%%%%%%%%%%%%%%%%

The CP on regular lattices with quenched infection rates has been 
studied by many authors (for a recent overview see \cite{Vojta}). 
First \cite{Noest} showed, using the Harris criterion \cite{harris74},
that spatially quenched disorder (frozen in space) changes the critical
behavior of the directed percolation for $D < 4$. Field theoretical 
RG~\cite{Janssen97b} found quenched disorder to be a marginal 
perturbation below $D < 4$ and the stable fixed point shifted to 
an unphysical region.
This means that spatially quenched disorder changes the critical behavior
of the directed percolation. This conclusion is supported by simulation 
results \cite{MoreiraDickman96,MF08,TFM09}.
In the sub-critical region they found GP, in which the time dependence is 
governed by non-universal power-laws, while in the active phase the 
relaxation of activity survival is algebraic.
A real-space RG study by \cite{HIV02} showed that in case of strong enough 
disorder the critical behavior is controlled by the infinite randomness 
fixed point and below $\lambda_c$ GP behavior emerges. Very recently
GP is reported in the five dimensional CP below the clean, mean-field 
critical point \cite{HoyosVojta}. 
 
Here I consider the one-dimensional SIS model with quenched disorder (QSIS), 
which exhibits $i\leftrightarrow j$ symmetry in the governing Eq.~(\ref{qmfsis}). 
First I investigated the case of uniformly distributed disorder, by putting 
symmetric weights, drawn from the distribution $w_{i,i+1} \in (0,1)$, on 
the edges connecting neighbors.

\begin{figure}
\includegraphics[height=5.5cm]{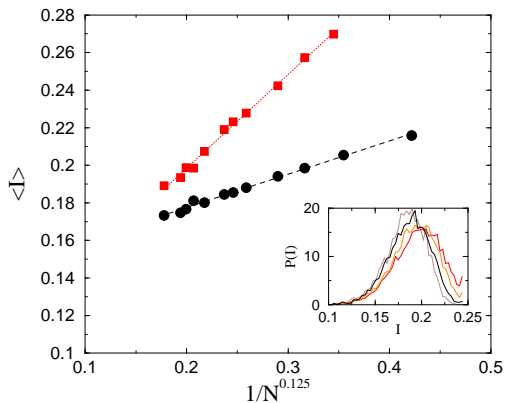}
\caption{\label{QSIS}(Color online) Finite size scaling of the IPR results 
of the one-dimensional QSIS model. Mean values of IPR for 
uniformly (bullets) and binary (squares) distributed disorder. 
Dashed line shows an extrapolation to $N\to \infty$ as: 
$0.13(1) + (0.17/N)^{0.1}$. 
Dotted line: $0.12(1) + (0.54/N)^{0.15}$. 
Inset: distributions of $I(N)$ in case of uniform distributed disorder
for various sizes. }
\end{figure}

The spectral analysis was done using the sparse matrix functions
of the software package OCTAVE \cite{OCT}. 
The largest eigenvalues and eigenvectors were determined and 
averaged over thousands of disorder realizations
for  $N= 10^3, ..., 5\times 10^5$. The probability distribution of
(\ref{IPR}): $P(I(N))$ is calculated by histogramming with the bin
size: $\delta I=0.001$.
As one can see on Figure~\ref{QSIS} the mean values of IPR remain finite 
and localization persists for any size. The $P(I(N))$ distributions do 
not smear, but shift to slower values by increasing the size.
In the $N\to\infty$ limit one can extrapolate the mean values 
${\bar P(I(N))}$ with a power-law, resulting in the asymptotic
value $I=0.168(2)$. 

The case of bimodal disorder distribution, where a fraction $q$ take 
a reduced value $r \lambda $, while the remaining fraction of the 
nodes take a value $(1-r) \lambda$:
\begin{equation}\label{bimodal}
p(\lambda_i) = (1-q) \delta[\lambda_i - (1-r)\lambda] + 
q \delta(\lambda_i - r \lambda) \
\end{equation}
has also been studied.
For $q=0.5$ only slow convergence of $I(N)$ could be observed, 
so I used a strong disorder distribution: $p=0.1$ and $r=0.9$.
In this case the IPR values are larger than for uniform distribution but
extrapolate roughly to the same  $I=0.125(15)$ value in the thermodynamic limit.

The finite size scaling of the largest eigenvalue determines the critical 
point within the QMF approximation $\lambda_c=1/y_M$. This extrapolates 
with a similar correction to scaling as for $I(N)$ to the value 
$\lambda_c = 0.548 + (0.35/N)^{0.27}$. Naturally, this value is 
much smaller than the true critical point of the model
due to the nature of approximations made.

%%%%%%%%%%%%%%%%%%%%%%%%%%%%%%%%%%%%%%%%%%%%%%%%%%%%%%%%%%%%%%%%%%%%%%%%%
%%%%%%%%%%%%%%%%%%%%%%%%%%%%%%%%%%%%%%%%%%%%%%%%%%%%%%%%%%%%%%%%%%%%%%%%

Thus the IPR, defined in the supercritical phase, predicts a localization 
in agreement with the known RR effects of CP in one-dimension.
Note, that for left-right asymmetric disorder, when sites interact with
their right or left neighbors, the localization disappears in the 
$N\to\infty$ limit in agreement with the recent results \cite{Robiuj}.

%%%%%% LIFS QSIS

\begin{figure}
\includegraphics[height=5.5cm]{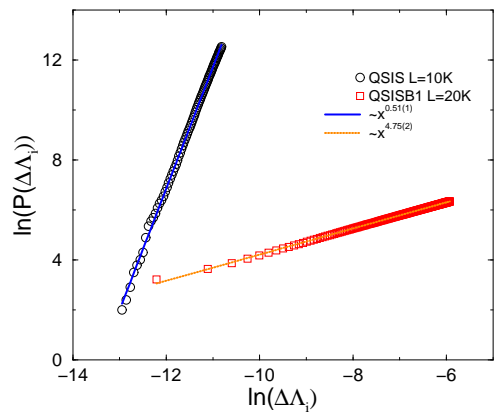}
\caption{\label{LQSIS10}(Color online)
Lifschitz tails in 1D QSIS models. Squares: tail distribution of the
$N = 2\times 10^4$ QSIS with bimodal random infection rates.
Circles: tail distribution of the $L=10^4$ QSIS with uniform random
distribution of infection rates. This curve is shifted by $\ln(10)$
both in $x$ and $y$ direction for better visibility.
Solid line shows a power-law fit $\sim \ln(\Delta\Lambda_i)^{4.75(5)}$, 
dotted line : power-law fit $\sim \ln(\Delta\Lambda_i)^{0.51(1)}$.}
\end{figure}

For the QSIS model the lower tail of the Laplacian has been determined
numerically for $N = 2\times 10^4$ . As Fig.~\ref{LQSIS10} shows
one can fit the tails with power-laws well. For uniform distribution
\begin{equation}
P(\Lambda) \sim \Lambda^{4.75} \ ,
\end{equation}
suggesting a GP behavior with decay law
\begin{equation}
\rho(t) \propto t^{-5.75 \lambda} 
\end{equation}
similarly for the known result of the CP.

%%%%%%%%%%%%%%%%%%%%%%%%%%%%%%%%%%%%%%%%%%%%%%%%%%%%%%%%%%%%%%%%%%%%%%%%%
%%%%%%%%%%%%%%%%%%%%%%%%%%%%%%%%%%%%%%%%%%%%%%%%%%%%%%%%%%%%%%%%%%%%%%%%

In conclusion I demonstrated here that even for this low dimensional model,
where dynamical fluctuations are relevant at the critical point
the effect of quenched disorder away from $\lambda_c$ can be well
described via the QMF approximation.

%%%%%%%%%%%%%%%%%%%%%%%%%%%%%%%%%%%%%%%%%%%%%%%%%%%%%%%%%%%%%%%%%%%%%%%%
\section{Localization transition on a generalized small world network model}
%%%%%%%%%%%%%%%%%%%%%%%%%%%%%%%%%%%%%%%%%%%%%%%%%%%%%%%%%%%%%%%%%%%%%%%%

In this section I show results of the QMF analysis done on networks, 
which exhibit purely topological disorder.
I analyzed a generalized {\it small-world} (GSW) network model
\cite{an,kleinberg,BB,Juhasz,Juhasz3}, which exhibits finite $D$, 
defined as follows. We add to a one-dimensional lattice (a ring) 
a set of long-range edges of 
arbitrary, unbounded, length. The probability that a pair of sites 
separated by the Euclidean distance $l$ is connected by an 
edge decays with $l$ as
\begin{equation}\label{pldist}
P(l)\simeq \beta l^{-s}
\end{equation} 
for large $l$ and amplitude $\beta$.
These networks interpolate between the quasi-one-dimensional network
($s=\infty$) and the mean-field limit ($s=0$). Recently, simulations 
of the CP provided numerical evidence for the emergence of GP in
$s\ge 2$ networks \cite{GPCNlong}.
When a quenched disorder added to the birth process rates a recent RG
study arrived to similar conclusions \cite{JK13}.

\begin{figure}
\includegraphics[height=5.5cm]{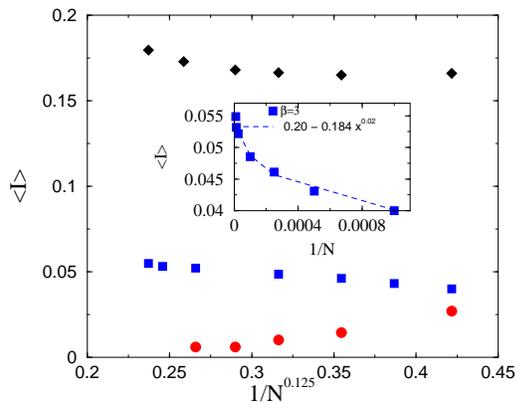}
\caption{\label{IPRBB}(Color online) Mean values of IPR of the SIS
model on GSW networks. For $s=2$, $\beta=0.1$ (diamonds) extrapolation 
to $N\to\infty$ shows the localization of the principal eigenvector. 
For $s=1$, $\beta=0.1$ (bullets) one observes a homogeneous steady 
state in the thermodynamic limit.
For $s=2$, $\beta=3$ (squares) a slow crossover to localization seems to emerge. 
On the main plot abscissa is rescaled to allow better visibility of
the finite size scaling. Inset: The crossover region magnified, rescaled 
and fitted with a power-law.}
\end{figure}

Here I show the finite size scaling results of $\langle I(N)\rangle$ 
defined on these GSW networks for sizes $N= 10^3, ...2\times 10^5$.
As Fig.~\ref{IPRBB} shows a clear localization occurs in the $s=2$ case 
with $\beta=0.1$ for $N\to\infty$.
For $s=2$ and large $\beta$, where the CP simulations and the RG analysis
were not completely conclusive, a slow crossover to localization can 
be concluded using an extrapolation to the data points: 
$I(N) = 0.20(2) - 0.18 (1/N)^{0.02}$  (see inset of Fig.~\ref{IPRBB}).
Here, the unusually small crossover exponent expresses the very slow 
change from small to large IPR in the infinite size limit.
%%%%%%%%%%%%%%%%%%%%%%%%%%%%%%%%%%%%%%%%%%%%%%%%%%%%%%%%%%%%%%%%%%%%%%%%%
%%%%%%%%%%%%%%%%%%%%%%%%%%%%%%%%%%%%%%%%%%%%%%%%%%%%%%%%%%%%%%%%%%%%%%%%
This result suggests that the GP of the SIS model may exist for any $\beta$ 
in case of marginal ($s=2$) GSW networks. 
In numerical simulations one should observe
GP regions of shrinking size, becoming invisible for large $\beta$-s.
Finally, for $s=1$ one observes a homogeneous steady state above the
critical point.

\begin{figure}
\includegraphics[height=5.5cm]{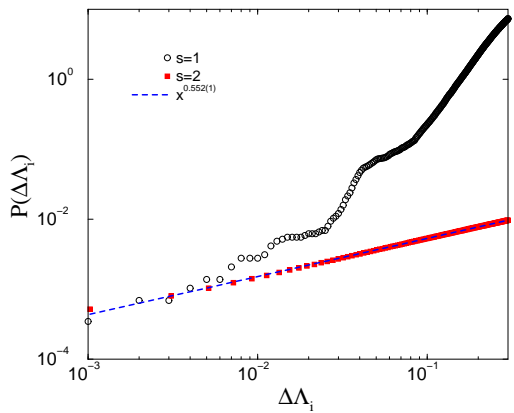}
\caption{\label{LsajeBBSISL}(Color online) Lifschitz tails on GSW 
graphs with $N=4\times 10^4$. Bullets: $s=1$, squares: $s=2$. 
Dashed line: power-law fitting: $\sim\Delta\Lambda_i^{0.55(1)}$.}
\end{figure}

%%%% LIF GSW
The $L_{ij}$ matrices have also been diagonalized for $N \le 4\times 10^4$ 
in case of $s=1$ and $s=2$ networks with $\beta=0.1$.
Dropping the trivial $\Lambda_1=0$ eigenvalues I calculated the
probability distribution of the smallest $500$ eigenvalues of the 
spectrum gap: $P(\Delta\Lambda_i) = P(\Lambda_i-\Lambda_2)$.
For the $N = 4\times 10^4$ networks the Lifschitz tail results are 
summarized on Fig.~\ref{LsajeBBSISL}. 
For $s=2$ a power-law tail emerges clearly, which can be fitted well using 
the least squares error method as $\sim\Delta\Lambda_i^{0.55(1)}$, 
in agreement with the expected GP behavior.
Contrary, at $s=1$ a deviation from power-law behavior can be observed 
on the log.-log. plot, the $P(\Delta\Lambda_i)$ curve grows faster than 
a simple power-law. Plotting $s=1$ curves on lin.-log. scale 
an exponential initial tail can be detected for $(\Delta\Lambda_i)<0.1$, 
slowing down later in a network size dependent way.

%%%%%%%%%%%%%%%%%%%%%%%%%%%%%%%%%%%%%%%%%%%%%%%%%%%%%%%%%%%%%%%%%%%%%%%%%
%%%%%%%%%%%%%%%%%%%%%%%%%%%%%%%%%%%%%%%%%%%%%%%%%%%%%%%%%%%%%%%%%%%%%%%%

%%%%%%%%%%%%%%%%%%%%%%%%%%%%%%%%%%%%%%%%%%%%%%%%%%%%%%%%%%%%%%%%%%%%%%%%
\section{Localization transition on scale-free networks}
%%%%%%%%%%%%%%%%%%%%%%%%%%%%%%%%%%%%%%%%%%%%%%%%%%%%%%%%%%%%%%%%%%%%%%%%

Up to know I showed agreement and success of the QMF-IPR method
by predicting the RR effects in agreement with the expectations. 
Now I point out some limitations.
Problems arise for example in case of SIS model on Barabasi-Albert (BA)
networks \cite{Barabasi:1999}. These networks are generated by a 
linear preferential attachment rule, starting from a small fully 
connected seed ($N_0$). At each time step $s$, a new vertex (labeled by $s$) 
with $m$ edges is added to the network and connected to an existing 
vertex $s'$ of degree  $k_{s'}$ with the probability
$\Pi_{s  \rightarrow s'} = k_{s'} /\sum_{s''= 1}^{s''<s} k_{s''}$.
By iterating the attachments for $N$ times one arrives to a graph
with $N+N_0$ nodes with an asymptotic SF degree distribution
$P(k) \simeq k^{-3}$.

\begin{figure}
\includegraphics[height=5.5cm]{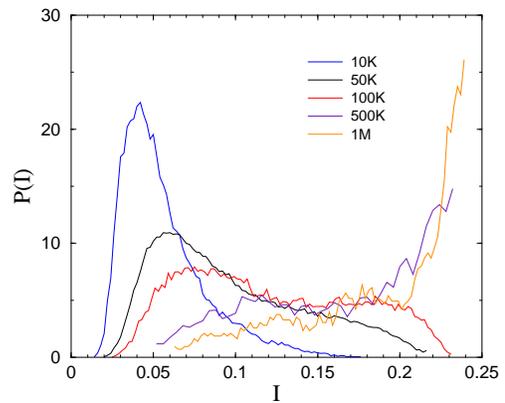}
\caption{\label{BA3IPR}(Color online) Probability distribution of IPR of 
the $m=3$ BA SIS model for sizes  $N=10^4$, $5\times10^4$ $10^5$, 
$5\times 10^5$ and $N=10^6$ (from left to right).}
\end{figure}

A previous work \cite{basiscikk} showed that for SIS models the IPR 
remained small in these networks, however uncertainties grew by increasing $N$. 
Now I compute IPR for a large number of disorder realizations for each
each $N$ and show that $P(I)$ distributions become wide as $N\to\infty$, 
with the appearance on an additional peak, besides the one at zero.
As shown on Fig.~\ref{BA3IPR} the second peak becomes dominant for 
$N\ge5\times10^5$, suggesting a crossover to localization in the infinite 
size limit. In this analysis $m=3$ BA networks with $N_0=5$ were used.

Earlier dynamical simulations of the CP did not show deviations from
the mean-field transition in case of $\gamma=3$ degree distribution, except
for BA trees, especially when certain weighting schemes were applied 
\cite{BAGPcikk}.
Note, that for SIS model $\lambda_c=0$ is expected in the  $N\to\infty$ 
limit, thus RR effects, if occur, could only slow down the relaxation 
towards the active state. 

%%%%%%%%%%%%%%%%%%%%%%%%%%%%%%%%%%%%%%%%%%%%%%%%%%%%%%%%%%%%%%%%%%%%%%%%%
%%%%%%%%%%%%%%%%%%%%%%%%%%%%%%%%%%%%%%%%%%%%%%%%%%%%%%%%%%%%%%%%%%%%%%%%

To investigate this further I considered the SIS on uncorrelated
configuration model (UCM) \cite{CBS05}, since one can control the degree 
distribution easily and these have been studied by various techniques. 
The UCMs were generated by the standard way.
In a set of $N$ vertexes one assigns to each vertex $k_i$ number 
of “stubs”, drawn from the probability distribution $P(k)$, with 
the $k_0 \le k_i < k_c$ and the $\mod(\sum_i k_i,2)=0$ constraints.
The network is completed by connecting pairs of these stubs
chosen randomly to form edges, respecting $k_i$ and avoiding self
or multiple connections. A minimum degree $k_0=2$ and a structural
cutoff $k_c=N^{1/2}$ was used to generate uncorrelated connected 
networks with probability one. The result of this construction
is a random network, whose degrees are distributed 
according to $P(k)$ without degree correlations.

\begin{figure}
\includegraphics[height=5.5cm]{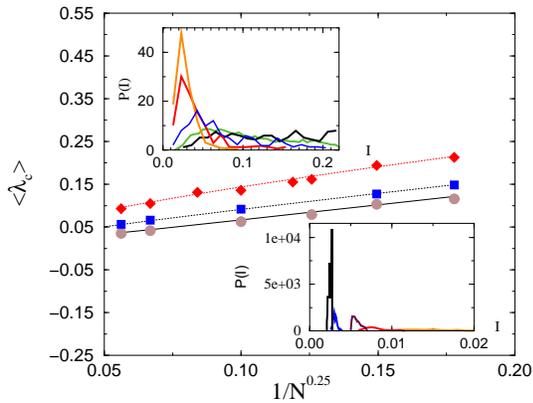}
\caption{\label{sajeUCM}(Color online) Finite size scaling of
QMF results of SIS on UCM networks. Mean value of $\lambda_c$ for 
$N= 10^3, ..., 2\times 10^5$ and for $\gamma=4$ (diamonds), $3$ (squares), 
$2.5$ (bullets). Lines correspond to power-law fits.
Right inset: $P(I)$-s for $\gamma=2.5$ with network sizes increasing 
from right to left curves. Left inset: $P(I)$-s for $\gamma=3$
with network sizes increasing from the left to right curves.}
\end{figure}

I generated the adjacency matrices for a large number of UCM 
graph realizations for degree distributions with $\gamma=4$, $3.5$, 
$3$, $2.8$, $2.5$ and performed the QMF analysis for sizes: $N=10^3$, 
$2\times10^3$, $10^4$, $5\times 10^4$, $10^5$, $2\times 10^5$.  
The estimated threshold values $\lambda_c = 1/y_M$ tend to zero 
in the $N\to\infty$ limit, in agreement with theoretical 
arguments for SIS: $\lambda_c \sim 1/N^{1/4}$ for $2.5 < \gamma\le 3$
and $\lambda_c \sim 1/N^{1/[2(\gamma-1)]}$ for $\gamma > 3$ \cite{Fer14,CP10}.
Power-law fits provided $\langle \lambda_c\rangle \sim 1/N^{0.25(1)}$
for $\gamma=2.5, 3$ and  $\langle \lambda_c\rangle \sim 1/N^{0.17(1)}$
for $\gamma = 4$ (see Fig.~\ref{sajeUCM}).

The probability distributions of IPR values are also calculated
and as the right inset of Fig.~\ref{sajeUCM} shows they converge 
to a sharp peak at $I=0$ for $\gamma=2.5$, $2.8$, while they smear, 
similarly as in case of BA, suggesting a localization
transition at $\gamma=3$ (see left inset of Fig.~\ref{sajeUCM}).
For $\gamma=4$, $3.5$ the peaks of $P(I(N))$ are localized around 
$I \simeq 0.25$.

%%%%%%%%%%%%%%%%%%%%%%%%%%%%%%%%%%%%%%%%%%%%%%%%%%%%%%%%%%%%%%%%%%%%%%%%%
%%%%%%%%%%%%%%%%%%%%%%%%%%%%%%%%%%%%%%%%%%%%%%%%%%%%%%%%%%%%%%%%%%%%%%%%

\begin{figure}
\includegraphics[height=5.5cm]{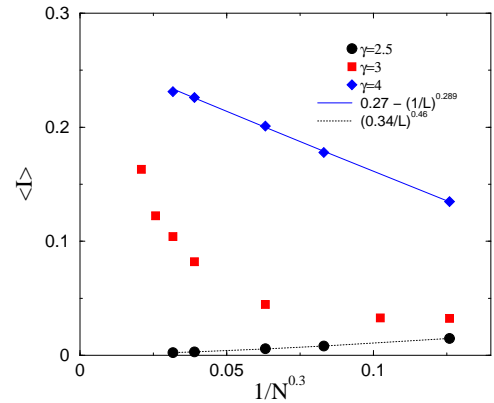}
\caption{\label{IPRUCM}(Color online) Mean values of IPR on UCM graphs
with $N= 10^3, ..., 2\times 10^5$. Rhombuses: $\gamma=4$ extrapolation
$N\to\infty$ results in $I=0.26(1)$ (localization). Bullets: $\gamma=2.5$. 
For $\gamma=3$ (squares) a crossover (a localization transition) emerges. }
\end{figure}

The mean value results of the IPR distributions are summarized on 
Fig.~\ref{IPRUCM}. 
For $\gamma=2.5$ networks $\lim_{N\to\infty} \langle I(N)\rangle=0$, 
thus no sign of localization appears. 
On the other hand, for $\gamma=4$ the mean
IPR remains finite and a localized network with $I(N) \to 0.26(1)$ 
can clearly be observed. Data are plotted on the $1/N^{0.3}$ scale, which
shows the leading order finite size scaling in the best way.
At $\gamma=3$ we can see a crossover towards eigenvector localization.
The distribution of $\langle I(N)\rangle$ is very wide here as in 
case of the BA graph.

%%%%%%%%%%%%%%%%%%%%%%%%%%%%%%%%%%%%%%%%%%%%%%%%%%%%%%%%%%%%%%%%%%%%%%%%%
%%%%%%%%%%%%%%%%%%%%%%%%%%%%%%%%%%%%%%%%%%%%%%%%%%%%%%%%%%%%%%%%%%%%%%%%

The coefficients of the expansion $a_1$, $a_2$ and $a_3$ in
Eq.~(\ref{expansion}) disappear as $\sim (1/N)$ in case of
$\gamma \ge 3$. On the other hand for $\gamma < 3$, $a_1$ decays 
slower than  $\sim (1/N)$, while $a_2$ and $a_3$ are roughly
zero, corresponding to a clear mean-field transition with $\beta=1$.
Such change has been observed in \cite{wbacikk,basiscikk}
in accordance with the emergence of RR effects.

\begin{figure}
\includegraphics[height=5.5cm]{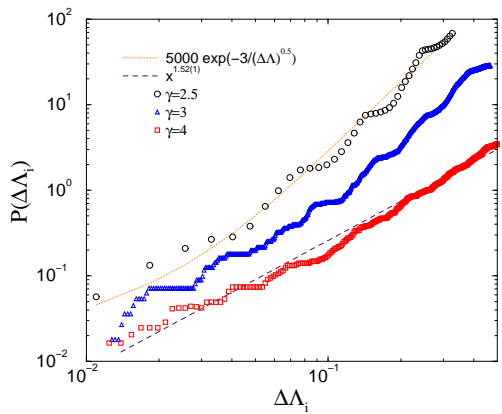}
\caption{\label{LsajeLUCML}(Color online) Lifschitz tails of SIS
on UCM graphs Bullets: $\gamma=4$, triangles: $\gamma=3$,
squares: $\gamma=2.5$.
Dashed line: power-law fitting with $\sim (\Delta\Lambda_i)^{1.52(1)}$.
Dotted line: least squares fitting with the stretched exponential form
$5000 \exp(-3/(\Delta\Lambda)^{0.5})$.}
\end{figure}

%%%%% LIFT
I have also studied the Lifschitz tail above and below the localization
transition in a similar way as before on UCM graphs with $N = 10^5$
nodes. The spectrum gap grows by decreasing $\gamma$ as:
$\Lambda_2=2.0166$ for $\gamma=4$, $\Lambda_2=2.0359$ for $\gamma=3$
and $\Lambda_2=2.0961$ for $\gamma=2.5$. This is in agreement with our
expectations, because larger gap means more entangled networks,
in which epidemic spreads quickly. The lowest $500$ eigenvalues are
calculated and histogrammed using bin sizes $\delta\Lambda=0.0001$,
following the drop of the $\Lambda_1=0$ eigenvalue. 
The $P(\Lambda_i)$ distributions
are shifted by $-\Lambda_2$ helping us to recognize possible power-laws
on log.-log. plots.
As Fig.~\ref{LsajeLUCML} shows, in the localized phase ($\gamma=4$)
a power-law distribution seems to emerge indeed, characterized by
$P(\Delta\Lambda_i) = 2.258(12) (\Delta\Lambda_i)^{1.52(1)}$.
On the other hand in the delocalized phase, for $\gamma=2.5$,
one can observe a faster than power-law behavior, which can be
fitted well with the stretched a exponential form: 
$5000 \exp(-3/(\Delta\Lambda)^{0.5})$, in agreement with the 
asymptotic of Eq.~(\ref{PER}), valid for uncorrelated random networks.
For comparison, a power-law fit assumption would lead to a large 
standard error of the regression coefficient: $\epsilon = 0.142$.
%%%%%%%%%%%%%%%%%%%%%%%%%%%%%%%%%%%%%%%%%%%%%%%%%%%%%%%%%%%%%%%%%%%%%%%%%
%%%%%%%%%%%%%%%%%%%%%%%%%%%%%%%%%%%%%%%%%%%%%%%%%%%%%%%%%%%%%%%%%%%%%%%%
Finally, at the $\gamma=3$ localization transition point, the tail
behavior at small $\Delta\Lambda$ deviates slightly away 
from a power-law, suggesting the lack of GP phase, in agreement
with the numerical simulations of \cite{BAGPcikk} done for the CP 
in BA networks. Assumption of a power-law fit form provides: $\epsilon = 0.018$.

Unfortunately the differences observed between the power-law and stretched
exponential tail behaviors are rather small. This is probably due to
the limitation of computing high precision $P(\Lambda)$ for large 
sizes. This puts a question mark on the applicability of the Lifschitz 
tail method in general.

%%%%%%%%%%%%%%%%%%%%%%%%%%%%%%%%%%%%%%%%%%%%%%%%%%%%%%%%%%%%%%%%%%%%%%%%
\section{Conclusions}
%%%%%%%%%%%%%%%%%%%%%%%%%%%%%%%%%%%%%%%%%%%%%%%%%%%%%%%%%%%%%%%%%%%%%%%%

Probability distributions of the inverse participation ratio have
been calculated in various network models exhibiting explicit or
topological heterogeneities. Careful finite size scaling analysis
pointed out the emergence of localization in generalized small world
and scale-free models. 
This method describes well the GP singularities
both in one-dimensional SIS with interaction disorder and in GSW-s 
with topological heterogeneity.
Localization, appearing in the active phase signals GP singularities
there. Former dynamical simulations in case of generalized small 
world networks \cite{PhysRevLett.105.128701,GPCNlong,MMNat} 
support this.
In infinite dimensional systems like ER graphs or BA networks GP-s with 
slow dynamics have been shown to appear only in weighted models 
\cite{PhysRevLett.105.128701,GPCNlong,BAGPcikk,wbacikk,Buo}.
In SF models with pure topological disorder the simulations
have been concentrated on the location of the critical point by
calculating stationary quantities \cite{FCP12,BCS13,Fer14,Ferrnew}
and visible GP effects have not been reported yet. Only loop-less
BA trees showed non-trivial phase transition by very extensive 
density decay simulations \cite{BAGPcikk}. This just corresponds 
to the localization point, thus one can expect more rare-region effects for 
$\gamma > 3$ SF networks. Preliminary numerical simulations indicate 
a time window, with a power-law like approach to the steady state.

On the other hand the lower spectral tail of the Laplacian describes
behavior in the inactive phase. In the linear approximation this is 
related to the dynamics of the order parameter.
The predictive power of the Lifschitz tail method has been investigated
and found qualitative agreement with the expectations. Finite systems 
exhibit spectral gaps, above which power-law tails were found, when GP 
behavior is expected. Fat tail distributions of the adjacency matrix of SF
were already shown in \cite{DGMS03}. This study suggests the existence
of SIS network models with fat-tailed Laplacians.
However, calculation of the Lifschitz tail numerically is a demanding task, 
not much easier than simulation of the time dependent order parameter. 
Furthermore, since QMF predicts $\lambda_c=0$ for SF and GSW networks
one cannot deeply be in the inactive phase of SIS, where the method is
expected to work in the thermodynamic limit.   

Application of these methods to SF networks results in a localization 
transition at $\gamma = 3$ both for correlated and uncorrelated graphs.
This is in agreement with the very recent simulation results, discussed 
in \cite{Ferrnew} and with the threshold, where the degree fluctuations 
$\langle k^2\rangle$ diverge in the HMF approximation \cite{PV01}
due to the strong heterogeneities. 
The localization in the active phase suggests dynamical RR effects 
for $\gamma > 3$, like in the models presented in \cite{LSN12}. 
However, for SIS, where $\lambda_c=0$ is expected to be in the thermodynamic 
limit, this implies a smeared phase transition, with an algebraic decaying 
density in a time window towards the active steady state value.
This scenario is feasible, because subspaces of an infinite dimensional
graph can be RR-s with arbitrary topological dimensions exhibiting 
phase transition at different $\lambda$-s, as suggested in \cite{BAGPcikk}.
According to \cite{CPN12} for large $\gamma$-s hubs sustain the epidemic 
processes instead of the the innermost, dense core, thus one may expect
that hubs play the role of RR-s here. 
In finite networks the smeared phase transition may also look like 
multiple phase transitions.

The success of the QMF method for describing GP behavior is demonstrated
here for SIS in basic network models. 
However, the linearization \cite{Monthus,KI11,JK13} and the complete 
neglection of dynamical fluctuations \cite{Fer14} warn for limitations 
on this relatively fast method, especially when the strong fluctuations 
override the localization effects.
The appearance of strong RR effects above the upper critical dimension 
\cite{HoyosVojta} supports, that QMF method is capable to predict exotic 
GP-s with off-critical, power-law singularities.

\section*{Acknowledgments}

I thank R. Juh\'asz for useful discussions and S. C. Ferreira, 
P. Van Mieghem for their comments. 
Support from the Hungarian research fund OTKA (Grant No. K109577) 
and the European Social Fund through project 
FuturICT.hu (grant no.: TAMOP-4.2.2.C-11/1/KONV-2012-0013) is acknowledged.


\begin{thebibliography}{50}%
\makeatletter
\providecommand \@ifxundefined [1]{%
 \@ifx{#1\undefined}
}%
\providecommand \@ifnum [1]{%
 \ifnum #1\expandafter \@firstoftwo
 \else \expandafter \@secondoftwo
 \fi
}%
\providecommand \@ifx [1]{%
 \ifx #1\expandafter \@firstoftwo
 \else \expandafter \@secondoftwo
 \fi
}%
\providecommand \natexlab [1]{#1}%
\providecommand \enquote  [1]{``#1''}%
\providecommand \bibnamefont  [1]{#1}%
\providecommand \bibfnamefont [1]{#1}%
\providecommand \citenamefont [1]{#1}%
\providecommand \href@noop [0]{\@secondoftwo}%
\providecommand \href [0]{\begingroup \@sanitize@url \@href}%
\providecommand \@href[1]{\@@startlink{#1}\@@href}%
\providecommand \@@href[1]{\endgroup#1\@@endlink}%
\providecommand \@sanitize@url [0]{\catcode `\\12\catcode `\$12\catcode
  `\&12\catcode `\#12\catcode `\^12\catcode `\_12\catcode `\%12\relax}%
\providecommand \@@startlink[1]{}%
\providecommand \@@endlink[0]{}%
\providecommand \url  [0]{\begingroup\@sanitize@url \@url }%
\providecommand \@url [1]{\endgroup\@href {#1}{\urlprefix }}%
\providecommand \urlprefix  [0]{URL }%
\providecommand \Eprint [0]{\href }%
\providecommand \doibase [0]{http://dx.doi.org/}%
\providecommand \selectlanguage [0]{\@gobble}%
\providecommand \bibinfo  [0]{\@secondoftwo}%
\providecommand \bibfield  [0]{\@secondoftwo}%
\providecommand \translation [1]{[#1]}%
\providecommand \BibitemOpen [0]{}%
\providecommand \bibitemStop [0]{}%
\providecommand \bibitemNoStop [0]{.\EOS\space}%
\providecommand \EOS [0]{\spacefactor3000\relax}%
\providecommand \BibitemShut  [1]{\csname bibitem#1\endcsname}%
\let\auto@bib@innerbib\@empty
%</preamble>
\bibitem{ABR02} R. Albert and A.-L. Barab\'asi, 
Rev. Mod. Phys. {\bf 74}, (2002) 47.
\bibitem{DFR02} S. N. Dorogovtsev and J. F. F. Mendes, {\it Evolution of
networks: From biological nets to the Internet and WWW},
(Oxford Univ. Press Oxford, 2003).
\bibitem{NR10} M. E. J. Newman, {\it Networks: An introduction},
(Oxford Univ. Press Oxford, 2010).
\bibitem [{\citenamefont {Harris}(1974{\natexlab{a}})}]{harris74}%
  \BibitemOpen
  \bibfield  {author} {\bibinfo {author} {\bibfnamefont {T.~E.}\ \bibnamefont
  {Harris}},\ }\href@noop {} {\bibfield  {journal} {\bibinfo  {journal} {Ann.
  Prob.}\ }\textbf {\bibinfo {volume} {2}},\ \bibinfo {pages} {969} (\bibinfo
  {year} {1974}{\natexlab{a}})}\BibitemShut {NoStop}
\bibitem [{\citenamefont {Liggett}(1985)}]{liggett1985ips}%
  \BibitemOpen
  \bibfield  {author} {\bibinfo {author} {\bibfnamefont {T.~M.}\ \bibnamefont
  {Liggett}},\ }\href@noop {} {\emph {\bibinfo {title} {{Interacting Particle
  Systems}}}}\ (\bibinfo  {publisher} {Springer-Verlag},\ \bibinfo {address}
  {New York},\ \bibinfo {year} {1985})\BibitemShut {NoStop}
\bibitem{pv04} R. Pastor-Satorras and A. Vespignani,
{\it Evolution and Structure of the Internet: A Statistical Physics Approach}
(Cambridge University, Cambridge, 2004).
\bibitem{MMNat} P. Moretti, M. A. Mu\~noz, Nature Communications {\bf 4},
(2013) 2521.
\bibitem{SIS} N. T. J. Bailey, {\it The Mathematical Theory of Infectious
Diseases} (Griffin, London, 1975), 2nd ed.; J. D. Murray, { Mathematical
Biology} (Springer-Verlag, Berlin, 1993).
\bibitem{BCS13} M. Bogu\~na, C. Castellano and R.  Pastor-Satorras,
Phys. Rev. Lett. {\bf 111} (2013) 068701.
\bibitem [{\citenamefont {Marro}\ and\ \citenamefont
  {Dickman}(1999)}]{DickMar}
  \BibitemOpen
  \bibfield  {author} {\bibinfo {author} {\bibfnamefont {J.}~\bibnamefont
  {Marro}}\ and\ \bibinfo {author} {\bibfnamefont {R.}~\bibnamefont
  {Dickman}},\ }\href@noop {} {\emph {\bibinfo {title} {{Nonequilibrium Phase
  Transitions in Lattice Models}}}}\ (\bibinfo  {publisher} {Cambridge
  University Press},\ \bibinfo {address} {Cambridge},\ \bibinfo {year}
  {1999})\BibitemShut {NoStop}%
\bibitem{rmp} G. \'Odor, Rev. Mod. Phys. {\bf 76} (2004) 663
\bibitem [{\citenamefont {\'Odor}(2008)}]{odorbook}
  \BibitemOpen
  \bibfield  {author} {\bibinfo {author} {\bibfnamefont {G.}~\bibnamefont
  {\'Odor}},\ }\href@noop {} {\emph {\bibinfo {title} {Universality in
  Nonequilibrium Lattice Systems}}}\ (\bibinfo  {publisher} {World
  Scientific},\ \bibinfo {address} {Singapore},\ \bibinfo {year}
  {2008})\BibitemShut {NoStop}
\bibitem{HH} M. Henkel, H. Hinrichsen and S. L\"ubeck, 
{\it Non-Equilibrium Phase Transitions vol. 1.}, Springer 2008
\bibitem [{\citenamefont {Vojta}(2006)}]{Vojta}%
  \BibitemOpen
  \bibfield  {author} {\bibinfo {author} {\bibfnamefont {T.}~\bibnamefont
  {Vojta}},\ }\href@noop {} {\bibfield  {journal} {\bibinfo  {journal} {Journal
  of Physics A: Mathematical and General}\ }\textbf {\bibinfo {volume} {39}},\
\bibinfo {pages} {R143} (\bibinfo {year} {2006})}\BibitemShut {NoStop}
\bibitem [{\citenamefont {Mu\~noz}\ \emph {et~al.}(2010)\citenamefont
  {Mu\~noz}, \citenamefont {Juh\'asz}, \citenamefont {Castellano},\ and\
  \citenamefont {\'Odor}}]{PhysRevLett.105.128701}%
  \BibitemOpen
  \bibfield  {author} {\bibinfo {author} {\bibfnamefont {M.~A.}\ \bibnamefont
  {Mu\~noz}}, \bibinfo {author} {\bibfnamefont {R.}~\bibnamefont {Juh\'asz}},
  \bibinfo {author} {\bibfnamefont {C.}~\bibnamefont {Castellano}}, \ and\
  \bibinfo {author} {\bibfnamefont {G.}~\bibnamefont {\'Odor}},\ }\href@noop {}
  {\bibfield  {journal} {\bibinfo  {journal} {Phys. Rev. Lett.}\ }\textbf
  {\bibinfo {volume} {105}},\ \bibinfo {pages} {128701} (\bibinfo {year}
  {2010})}\BibitemShut {NoStop}%
\bibitem [{\citenamefont {Griffiths}(1969)}]{Griffiths}%
  \BibitemOpen
  \bibfield  {author} {\bibinfo {author} {\bibfnamefont {R.~B.}\ \bibnamefont
  {Griffiths}},\ }\href@noop {} {\bibfield  {journal} {\bibinfo  {journal}
  {Phys. Rev. Lett.}\ }\textbf {\bibinfo {volume} {23}},\ \bibinfo {pages} {17}
  (\bibinfo {year} {1969})}\BibitemShut {NoStop}%
\bibitem{Johnson} S. Johnson, J. J. Torres, and J. Marro, PLoS ONE
{\bf 8(1)}: e50276 (2013)
\bibitem{PV01} R.  Pastor-Satorras and A. Vesignani,
Phys. Rev. Lett. {\bf 86} (2001) 3200.
\bibitem{KK11} M. Karsai et al. Phys. Rev. E {\bf 83} (2011) 025102(R)
\bibitem{Chialvo} A. Haimovici, E. Tagliazucchi, P. Balenzuela and 
D. R. Chialvo, Phys. Rev. Lett. {\bf 110}, 178101 (2013).
\bibitem{burstcikk} G. \'Odor, Phys. Rev. E {\bf 89}, 042102 (2014)
\bibitem{IMrev} F. Igl\'oi, C. Monthus, Phys. Rep. {\bf 412}, 277-431, (2005)
\bibitem{KK93} B. Kramer and A. MacKinnon, Rep. Prog. Phys. {\bf 56}, 1469
(1993).
\bibitem [{\citenamefont {{Bogu\~{n}\'{a}}}\ \emph {et~al.}(2009)\citenamefont
  {{Bogu\~{n}\'{a}}}, \citenamefont {Castellano},\ and\ \citenamefont
  {Pastor-Satorras}}]{boguna09:_langev}%
  \BibitemOpen
  \bibfield  {author} {\bibinfo {author} {\bibfnamefont {M.}~\bibnamefont
  {{Bogu\~{n}\'{a}}}}, \bibinfo {author} {\bibfnamefont {C.}~\bibnamefont
  {Castellano}}, \ and\ \bibinfo {author} {\bibfnamefont {R.}~\bibnamefont
  {Pastor-Satorras}},\ }\href@noop {} {\bibfield  {journal} {\bibinfo
  {journal} {Phys. Rev. E}\ }\textbf {\bibinfo {volume} {79}},\ \bibinfo
  {pages} {036110} (\bibinfo {year} {2009})}\BibitemShut {NoStop}%
\bibitem{FCP12} S. C. Ferreira, C. Castellano, R. Pastor-Satorras,
  Phys. Rev. E {\bf 86}, 041125 (2012)
\bibitem{Fer14} Ang\'elica S. Mata, Ronan S. Ferreira, Silvio C. Ferreira,
New J. Phys. {\bf 16} (2014) 053006
\bibitem{CWW08} D. Chakrabarti, Y. Wang, C. Wang, J. Leskovec, and
C. Faloutsos, ACM Trans. Inf. Syst. Secur. {\bf 10}, 1 (2008).
\bibitem{Mieg09} P. Van Mieghem, J. Omic, and R. Kooij, IEEE ACM T.
Network. {\bf 17}, 1 (2009).
\bibitem{CP10} C. Castellano and R.  Pastor-Satorras,
Phys. Rev. Lett. {\bf 105} (2010) 218701.
\bibitem{GDOM12} A. V. Goltsev, S. N. Dorogovtsev, J. G. Oliveira,
        and J. F. F.  Mendes, Phys. Rev. Lett. {\bf 109}, 128702 (2012)
\bibitem{wbacikk} G.\'Odor, Phys. Rev. E {\bf 87}, (2013) 042132.
\bibitem{basiscikk} G.\'Odor, Phys. Rev. E {\bf 88}, (2013) 032109.
\bibitem{Monthus} C. Monthus, T. Garel, J. Phys. A: Math. Theor. {\bf 44} (2011) 085001.
\bibitem{KI11}I. A. Kov\'acs and F. Igl\'oi, J. Phys.: Condens. Matter {\bf 23}, 404204 (2011).
\bibitem{JK13}R. Juh\'asz and I. A. Kov\'acs, J. Stat. Mech. (2013) P06003.
\bibitem{Matj} P. Villegas, P. Moretti, M. A. Mu\~noz, Sci. Rep. {\bf 4}, (2014), 5990
\bibitem{Nus91} A. N\"usser, J. Phys. A, {\bf 24} (1991) 2355.
\bibitem{KKM} O. Khorunzhiy, W. Kirsch, and P. M\"ller, Ann. Appl. Probab.
{\bf 16} (2006), 295-309.
\bibitem{Chung} F. R. K. Chung (1997). {\it Spectral Graph Theory}, 
Amer. Math. Soc., Providence, RI.
\bibitem{Miegbook} P. V. Mieghem, {\it Graph Spectra for Complex Networks},
(Cambridge University Press, Cambridge 2011).  
\bibitem{SDM08} A. N. Samukhin, S. N. Dorogovtsev and J. F. F. Mendes,
PRB {\bf 77}, 036115 (2008).
\bibitem{TLNG05} S. N. Taraskin et al, Phys. Rev. E. {\bf 7}, 016111
(2005).
\bibitem{BC} M. Barahona and L. M. Pecora, Phys. Rev. Lett {\bf 89} (2002) 054101
\bibitem{DNM06} L. Donetti, F. Neri and M. Mu\~noz, J. of Stat. Mech (2006) P08007.
\bibitem{Mieg12} P. Van Mieghem, Eur. Phys. Lett. {\bf 97}, 48004 (2012).
\bibitem{MO08} P. Van Mieghem and J. Omic, Delft University of Technology, 
report2008081, arxiv.org:1306.2588.
\bibitem{M11} P. Van Mieghem, Computing {\bf 93}, (2011) 147-169.
\bibitem{Noest} A. J. Noest, Phys. Rev. Lett. {\bf 57}, 90 (1986);
A. J. Noest, Phys. Rev. B {\bf 38}, 2715 (1988).
\bibitem{Janssen97b} H.~K. Janssen, Phys. Rev. E {\bf 55}, 6253 (1997).
\bibitem{MoreiraDickman96} A.~G. Moreira and R.~Dickman,
Phys. Rev. E {\bf 54}, R3090 (1996).
\bibitem{MF08} M. M. de Oliveira and S. C. Ferreira, JSTAT P11001 (2008)
\bibitem{TFM09} T. Vojta, A. Farquhar, and J. Mast, Phys. Rev. E {\bf 79}, 011111 (2009).
\bibitem{HIV02} J. Hooyberghs, F. Igl\'oi and C. Vanderzande,
Phys. Rev. Lett., {\bf 90}, 100601 (2003)
\bibitem{OCT} http://www.gnu.org/software/octave
\bibitem{an} M. Aizenman and C.M. Newman, Commun. Math. Phys. 
{\bf 107}, 611 (1986).
\bibitem{kleinberg}  J.M. Kleinberg, Nature {\bf 406}, 845 (2000).
\bibitem{BB}I. Benjamini and N. Berger, Rand. Struct. Alg. {\bf 19}, 102 (2001).
\bibitem [{\citenamefont {Barab{\'a}si}\ and\ \citenamefont
  {Albert}(1999)}]{Barabasi:1999}%
  \BibitemOpen
  \bibfield  {author} {\bibinfo {author} {\bibfnamefont {A.-L.}\ \bibnamefont
  {Barab{\'a}si}}\ and\ \bibinfo {author} {\bibfnamefont {R.}~\bibnamefont
  {Albert}},\ }\href@noop {} {\bibfield  {journal} {\bibinfo  {journal}
  {Science}\ }\textbf {\bibinfo {volume} {286}},\ \bibinfo {pages} {509}
  (\bibinfo {year} {1999})}\BibitemShut {NoStop}
\bibitem{Juhasz} R. Juh\'asz, Phys. Rev. E {\bf 78}, 066106 (2008).
\bibitem{Juhasz3} R. Juh\'asz, G. \'Odor, Phys. Rev. E  {\bf 80}, 041123 (2009)
\bibitem{GPCNlong} R. Juh\'asz, G. \'Odor, C. Castellano and M. Mu\~noz,
Phys. Rev. E {\bf 85}, 066125 (2012)
\bibitem{BAGPcikk} G.\'Odor and R. Pastor-Satorras, 
Phys. Rev. E {\bf 86}, (2012) 026117.
\bibitem{HoyosVojta} T. Vojta, J. A. Hoyos, Phys. Rev. Lett. {\bf 112}, 075702 (2014);
T. Vojta, J. Igo, J. A. Hoyos,  arXiv:1405.4337
\bibitem{Robiuj} R. Juh\'asz, J. Stat. Mech. (2013) P10023
\bibitem{ER} P. Erd\H os and A. R\'enyi, A. (1959) Publ. Math. {\bf 6}, 290–291.
\bibitem{CBS05} M. Catanzaro, M. Bogu\~na and R. Pastor-Satorras,
Phys. Rev. E {\bf 71} (2005) 027103.
\bibitem{DGMS03} S. N. Dorogovtsev, A. V. Goltsev, J. F. F. Mendes,
Phys. Rev. E {\bf 68}, 046109.
\bibitem{Ferrnew} A. S. Mata, S. C. Ferreira, arXiv:1403.6670
\bibitem{LSN12} H. K. Lee, P.-S. Shim and J. D. Noh,
Phys. Rev. E {\bf 87}, 062812 (2013)
\bibitem{CPN12} C. Castellano and R. Pastor-Satorras, Sci. Rep. 2 (2012) 371.
\bibitem{Buo} C. Buono, F. Vazquez, P. A. Macri, L. A. Braunstein,
    Phys. Rev. E {\bf 88}, 022813 (2013)
\end{thebibliography}
\end{document}